\documentclass[11pt,twoside]{article}
\usepackage{asp2004}
\usepackage{psfig}
\usepackage{epsf}
\usepackage{graphics}
\usepackage{lscape}
\def\edcomment#1{\iffalse\marginpar{\raggedright\sl#1\/}\else\relax\fi}
\reversemarginpar

\begin{document}

\title{Probing the Type Ia environment with Light Echoes}
\author{F. Patat}
\affil{E.S.O. K. Schwarzschild Str.2 - 85748 - Garching b. M\"unchen - Germany}

\begin{abstract}
In general, Light Echoes (LE) are beautiful, rather academical and
therefore unavoidably useless phenomena. In some cases, however, they
can give interesting information about the environment surrounding the
exploding star.  After giving a brief introduction to the subject, I
describe its application to the case of Type Ia Supernovae and discuss
the implications for progenitors and their location within the host
galaxies.
\end{abstract}
\thispagestyle{plain}

\section{Introduction}
\label{sec:intro}

The idea about a possible connection between the characteristics shown
by Type Ia Supernovae (hereafter SNe~Ia) and the properties of their
host galaxies has been around for quite a while.  Several authors have
pointed out that the observed features of Ia's, such as intrinsic
luminosity, colour, decline rate, expansion velocity and so on, appear
to be related to the morphological type of the host galaxy
(\citealt{flip89}, \citealt{branchvdb}, \citealt{vdb},
\citealt{hamuy96}, \citealt{hamuy00}, \citealt{howell}; see also the
contributions by Mannucci, Della Valle, Petrosian and Garnavich in
these proceedings). Since these objects represent a fundamental tool
in Cosmology, it is clear that a full understanding of the underlying
physics is mandatory in order to exclude possible biases when one is
to disentangle between different cosmological scenarios.

In this framework, recognizing the existence of Ia sub-classes is a
fundamental step. In this respect, a milestone in the SN history is
year 1991, when two extreme objects were discovered, i.e.  SN1991T
(\citealt{flip92a}) and SN1991bg (\citealt{flip92b}). The former was
an intrinsically blue, slow declining and spectroscopically peculiar
event, while the latter was intrinsically red, fast declining and also
showing some spectral peculiarities.  From that time on, several other
objects sharing the characteristics of one or the other event were
discovered, indicating that these deviations from the standard Ia
were, after all, not so rare. Of course, one of the most important
issues which were generated by the discovery of such theme variations
concerned the explosion mechanism and, in turn, the progenitor's
nature.

The growing evidences produced by the observations in the last ten
years have clearly demonstrated that the sub-luminous events
(1991bg-like) are preferentially found in early type galaxies (E/S0),
while the super-luminous ones (1991T-like) tend to occur in spirals
(Sbc or later). This has an immediate consequence on the progenitors,
in the sense that sub-luminous events appear to arise from an old
population while super-luminous ones would rather occur in
star-forming environments and therefore would be associated with a
younger population. This important topic has been discussed by
\citet{howell}, to which I refer the reader for a more detailed
review.  What is important to emphasize here is that 1991T-like events
tend to be associated with young environments and are, therefore, the
most promising candidates for the study of Light Echoes (LEs). Or, in
turn, if LEs are detected around such kind of SNe, this would
strengthen their association with sites of relatively recent star
formation.

\section{Known Light Echoes in SNe~Ia}
\label{sec:known}

Due to the typical number density of dust particles which are
responsible for the light scattering, LEs are expected to have an
integrated brightness about ten magnitudes fainter than the SN at
maximum (see for example \citealt{sparks94}). For this reason, a SN Ia
in the Virgo cluster is supposed to produce, if any, an echo at a
magnitude V$\sim$21.0. This has the simple consequence that it is much
easier to observe such a phenomenon in a Ia than in any other SN type,
due to its high intrinsic luminosity. As a matter of fact, only four
cases of {\it scattered} LEs are known: the SNe Ia 1991T
(\citealt{schmidt}, \citealt{sparks99}) and 1998bu (\citealt{capp}), and 
the type II SN1987A (\citealt{xu95} and references therein) and 1993J
(\citealt{sugerman}). As expected, the LE detections for the two
core-collapse events occurred in nearby galaxies: LMC (d=50 kpc) and
M81 (d=3.6 Mpc) respectively.

The first case of a LE in a Ia (1991T) seems to confirm the scenario
outlined in the Introduction, in the sense that the SN was
over-luminous and the host galaxy (NGC4527) is an Sbc and also a
liner. Slightly less convincing is the other known case (1998bu),
since the galaxy (NGC3368) is both an Sbc and a liner, but the SN is
not spectroscopically peculiar. The only characteristic in common with
SN1991T is its decline rate $\Delta m_{15}$, which is lower than
average, even though not so extreme as in the case of
1991T. Nevertheless, the HST observations by \cite{garnavich} show
that a significant amount of dust must be present within 10 pc from the
SN.  Of course no statistically significant conclusion can be drawn
from such a small sample, which definitely needs to be enlarged. For
this reason, during the past years, I have been looking for new cases,
the most promising of which was represented by SN1998es in
NGC~632. This SN, in fact, was classified as a 1991T-like by
\citet{jha}, who also noticed that the parent galaxy was an S0,
hosting a nuclear starburst \citep{pogge}. Moreover, the SN was found
to be projected very close to a star forming region and to be affected
by a strong reddening, which all together made SN~1998es a very good
candidate for a LE study.

The host galaxy was imaged with the ESO-VLT at almost three years
from the explosion. While I will give the details in a forthcoming
paper, here I can anticipate that no LE was detected, down to a
limiting magnitude of $V\sim$25. This allows one to definitely exclude
the presence of a 1991T-like LE, which would have been clearly
detectable by the VLT observations. Since the two objects suffered a
similar extinction, the only plausible explanation is that in 1998es
the dust was confined at a larger distance from the SN, at least a
factor 10 farther than in 1991T. Therefore, at least for this object,
the conjecture discussed in Sec.~\ref{sec:intro} is not confirmed.  As
a matter of fact, as a result of a systematic LE search including 64
historical SNe, \citet{boffi} have reported 16 possible candidates,
only one of which is a genuine Ia, i.e. SN~1989B (but see also
\citealt{milne}).  Therefore, one may first inquire why only two
events have been detected and immediately conclude that this is simply
because in the vast majority of the cases there is not enough, and/or
not close enough dust around Ia's. Not an unexpected conclusion for
supposedly long-lived and small mass progenitors.
 
Nevertheless, a consideration needs to be done. In the case of ground
based observations, I must notice that there are only a few Ia's
observed at more than one year past maximum. In fact, this has always
been a problem, both due to their faintness and to the presence of the
host galaxy background.  Therefore, it is difficult to give a final
answer on the basis of the exceedingly small list of LE detections and
we will probably have to wait a bit more in order to have a
statistically significant sample.  This is even more true if only
over-luminous SNe are associated with dusty regions.

\section{The Light Echo Phenomenon}
\label{sec:le}

Starting with the pioneering work by \cite{couderc}, the problem has
been addressed by several authors (see for example \citealt{dwek},
\citealt{chev},  \citealt{schae}, \citealt{emmering}, \citealt{sparks94}, 
\citealt{xu94} and \citealt{sugerman03}). The interested
reader can refer to these publications for a detailed description of
the phenomenon, while here I will give only a brief introduction.

The definition of LE is borrowed from the equivalent effect one can
easily experience with acoustic waves. A sound emitted by a source can
be reflected by the environment and reach the listener at different
times.  If the time delay is larger than the input sound duration one
has an {\it echo}, while in the other cases one should rather talk
about a {\it reverb}. If the environment is complex, the resulting
signal will be the superposition of the input transient (which in the
astronomical context corresponds to the source burst) and a large
number of delayed and filtered signals. In mathematical terms this is
described by the convolution of the signal with the impulse response
function (IRF), which contains the geometrical and physical properties
of the environment (see for example \citealt{spjut} for a good
introduction to this subject).  In the parallel astrophysical case,
the ingredients of the IRF are the density distribution of dust
particles, the scattering efficiency as a function of wavelength
(i.e. the extinction law), the dust albedo and the scattering
efficiency as a function of scattering angle (i.e. the scattering
phase function).

In the simplest case, the problem is solved in the so-called {\it
single scattering} approximation, which assumes that once a photon is
scattered by the dust, it escapes the system with no further
interaction. As already shown by \cite{chev}, this assumption holds
when the dust optical depth is low. When the optical depth grows,
multiple scattering becomes relevant and it produces quite interesting
effects. For a detailed description the reader is referred to
\cite{patat04}, where the Monte Carlo treatment, the ingredients and
the results are thoroughly discussed.

Due to the short duration of a Type Ia burst, the apparent LE is
approximately confined within a thin paraboloidal shell, with the SN
in its focus. As a consequence, at any given time, the observed
properties of the convolved signal reflect the geometrical structure
and the physical properties of a well defined portion of the dusty
environment.  Therefore, the hope behind this kind of analysis is that
a LE can be used as a tomographic probe. And this is indeed the case
when the echo is resolved, as it has been shown for SN~1987A by
Crotts, Xu and collaborators (see \citealt{xu95} and references
therein).  The problem becomes ill posed when the LE is unresolved,
since the simulations show that one can produce pretty similar light
curves and spectra with different dust distributions. This is because
there is a partial degeneracy in the dust-density/dust-distance plane,
which leaves one with the question whether there are lots of dust far
from the SN or a small amount of dust close to the SN.

\section{Applications to SNe~Ia}

As an example application to the study of SNe~Ia environments, I will
discuss here the simple case of an event occurring in a face-on dusty
disk. The density profile of such a system can be modeled using the
typical double exponential formulation, which is parameterized through
$R_d$ and $Z_d$, the characteristic radial and vertical scales for the
dust distribution.  In the case of a spiral galaxy, typical values are
$R_d$=4.0 kpc and $Z_d$=0.14 kpc. The central density, $n_0$, is
constrained by $\tau(0)$, i.e. the central optical depth of the disk
seen face-on.  Imposing a typical value $\tau_V(0)$=1 and for a
$R_V$=3.1 Milky Way dust mixture, $n_0$ turns out to be 2.3
cm$^{-3}$. Now, if we place the SN at 6 kpc from the center and at 0.1
kpc above the galactic plane (which implies an optical depth
$\tau_V$=0.06 for the SN), the resulting LE is about 10 mag fainter than
the SN at maximum, and shows a very low luminosity decline rate.

In conclusion, under rather normal conditions, a Type Ia SN exploding
in the disk of a spiral should always produce an observable LE,
without the SN being heavily reddened.  Of course, placing the SN in
the inner parts of the disk would increase the echo luminosity and the
extinction suffered by the SN itself. For example, leaving all the
other parameters unchanged and placing the SN on the galactic plane
would enhance the LE by 0.7 mag, while the optical depth would grow to
$\tau_V\sim$0.1, which is anyway still a rather low value.  Since
there are good reasons to believe that $1\leq\tau_V(0)\leq 5$, a Type
Ia within 1-2 dust scale heights should always produce an observable
LE, unless it is located very far from the galactic center. As a
conclusion, it seems that SNe~Ia tend to explode far from the host
galaxy disk, at vertical distances from the galactic plane that are
significantly larger than the dust height scale $Z_d$, otherwise they
would always produce detectable LEs.

\begin{figure}
\plotfiddle{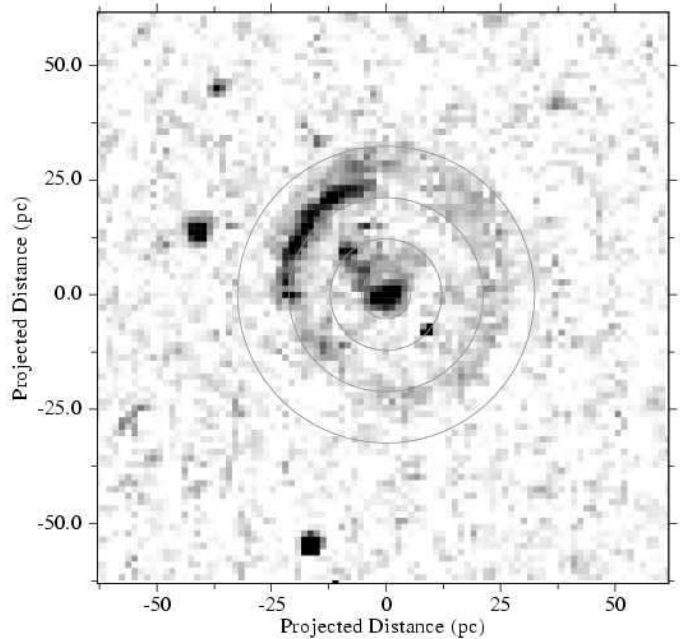}{7.7cm}{-90}{48}{48}{-185}{255}
\caption{\label{fig:98bu}Image of SN~1998bu obtained on 2003-04-23
($\sim$4.9 yrs after the explosion) with the High Resolution Camera of
ACS (F435W), mounted on board of the HST. The projected scale ($\sim$
1.3 pc pixel$^{-1}$) corresponds to a host galaxy distance of 10.5
Mpc. The circles trace the loci at constant radial distance from the
SN and were placed at 10, 50, 150 and 350 pc respectively.}
\end{figure}

As I have mentioned before, remarkable exceptions are SNe 1991T and
1998bu, whose LEs have been resolved by HST (see \citealt{sparks99}
and \citealt{garnavich}). Due to the smaller distance of its host
galaxy, the case of 1998bu is particularly interesting. The most
recent HST-ACS images available show a very clear ring-like structure,
as originally noticed by Garnavich and collaborators from the WFPC2
images taken about two years after the SN had exploded. This is
clearly visible in Fig.~\ref{fig:98bu}, where I have also plotted the
projected distance scale from the resolved central knot, whose
centroid practically coincides with the SN position. Overimposed are
also a number of circles, which define the loci of constant distance
$r$ from the SN, derived using the LE paraboloid equation (see for
example
\citealt{patat04}). Since background dust would appear at projected
distances smaller than $ct$ (which here corresponds to about 1.5 pc),
all the resolved structures visible in Fig.~\ref{fig:98bu} are
generated by foreground material, which extends to at least 300 pc
from the SN. In general, the LE can be roughly subdivided into 3
regions: a) an external ring, b) a central resolved disk and c) a
filamentary structure extending from the center to the strong density
enhancement visible at about $r$=180 pc.  Region a) is most likely
produced by dust confined within a non-planar and sheet-like cloud,
strongly inclined with respect to the line of sight and highly
inhomogeneous. The dust illuminated at this epoch and belonging to
this structure is placed at distances which range between 120 and 300
pc from the SN. Even taking into account the artificial broadening
effect produced by the finite duration of the input SN flash and the
instrumental spatial resolution, the maximum sheet thickness must be
larger than 100 pc. But the most interesting aspect of this LE, in
connection to our original problem, is the detection of dust
relatively close to the SN. The original finding by
\cite{garnavich} is confirmed by the latest HST observations presented
here, which reveal details on even closer regions: dust is certainly
present at $r<$10 pc. A sufficiently small distance to claim that the
SN progenitor was associated to the dust cloud.  A thorough and more
quantitative analysis of these and other HST archival data, including
polarimetry, is in progress.  This will provide a detailed study of
the LE evolution with time, a probe for the SN environment at
different locations and a tool to get dust density estimates.

Notwithstanding the difficulties which are intrinsic to the method, I
still think LEs can give us some insights on the SNe~Ia explosion
environment. 

Observers, watch out the late phases of your Ia's!

\vspace{5mm}

{\it Bench\'e l'astronomia nel corso di molti secoli abbia fatto gran
progressi nell'investigar la constituzione e i movimenti de i corpi
celesti non per\`o \`e ella sin qui arrivata a segno tale che
moltissime cose non restino indecise, e forse ancora molt'altre
occulte.\footnote{Even though astronomy during many a century has made
great progresses in investigating the constitution and motion of
celestial bodies, it has not yet reached such a level that many things
do not remain undecided, and perhaps many others still unknown.}}
\begin{flushright}\vspace{-3mm}Galilei (1632)\end{flushright}

\end{document}